# Random lasing actions in self-assembled perovskite nanoparticles

Shuai Liu[1], Wenzhao Sun[1], Jiankai Li[1], Zhiyuan Gu[1], Kaiyang Wang[1], Shumin Xiao[2, †], Qinghai Song[1, *]

1. Integrated Nanoscience Lab, School of Electrical and Information Engineering, Harbin Institute of Technoloy, Shenzhen, China, 518055
2. Integrated Nanoscience Lab, School of Material Science and Engineering, Harbin Institute of Technoloy, Shenzhen, China, 518055

† shuminxiao@gmail.com

* qinghai.song@hitsz.edu.cn

**Abstract**：

Solution-based perovskite nanoparticles have been intensively studied in past few years due to their applications in both photovoltaic and optoelectronic devices. Here, based on the common ground between the solution-based perovskite and random lasers, we have studied the mirrorless lasing actions in self-assembled perovskite nanoparticles. After the synthesis from solution, discrete lasing peaks have been observed from the optically pumped perovskites without any well-defined cavity boundaries. The obtained quality (Q) factors and thresholds of random lasers are around 500 and 60 μJ/cm$^2$, respectively. Both values are comparable to the conventional perovskite microdisk lasers with polygon shaped cavity boundaries. From the corresponding studies on laser spectra and fluorescence microscope images, the lasing actions are considered as random lasers that are generated by strong multiple scattering in random gain media. In additional to conventional single-photon excitation, due to the strong nonlinear effects of perovskites, two-photon pumped random lasers have also been demonstrated for the first time. We believe this research will find its potential applications in low-cost coherent light sources and biomedical detection.

# Introduction:

Very recently, organic based lead halide perovskites have received considerable research attentions due to their high potentials in cost-effective optoelectronics and photovoltaic devices [1-13]. Compared with the other solution-cast semiconductor processed at similar low temperature, perovskite films have much larger optical absorption coefficient, optimal bandgap, small defect density, and long electron/hole diffusion length [1, 2]. Based on these intrinsic advantages, the record power conversion efficiency (20.1%) has been recently obtained from solution-based perovskite thin film [3]. Meanwhile, due to the slow Auger recombination, perovskite films have also been pointed out to exhibit excellent coherent light emission properties [4]. Optically pumped micro- and nano- lasers have been demonstrated and the estimated quantum yields even approach 100% [5-8]. Up to now, the perovskite lasers have been well studied in single crystalline nanoplates with regular cavity shapes, e.g. square [5], hexagon [6], and nanowires [7, 8, 12]. The lasing actions in irregular and nanosized perovskites, which have the potentials to be infiltrated into biological tissues and to serve in pinpoint detection, have been rarely explored yet. While some groups have reported randomly distributed laser peaks, the underlying mechanisms for random lasing actions have not been discussed [13, 14].

Unlike the conventional well-defined microplates [5-8], the irregular nanoparticles usually self-assemble and form disordered gain media (see Fig. 1(d) for example). Therefore, another kind of mirrorless laser, random laser [15], can be considered. Soon after the first observations by Lawandy et al. [16] and Cao et al. [17], random lasers have been widely observed in many kinds of materials [18-23]. Most of the lasing behaviors in weak and strong scattering media have been well explained [24] and thus the research attentions on random lasers gradually shift to practical applications due to their intrinsic advantages in low cost and simple preparation. A number of practical applications have been proposed and experimentally realized, e.g illumination light sources for speckle-free laser imaging [25], low-threshold directional emission of random laser [26], biomedical diagnosis [27-29]. Comparing the random lasers and the solution processed perovskites, it is interesting to see that they have the common ground in low-cost, easy fabrication, and potential applications. In this sense, the combination of perovskites and random lasers shall have very bright future in practical applications. Here, by exploring the lasing actions and corresponding fluorescent microscope images, the random lasing actions in disordered perovskites nanocrystals have been studied.

## Results and Discussions

Similar to the recent reports [5, 7, 9-11], our perovskite nanoparticles were synthesized using a simple one-step solution self-assemble method. The $CH_3NH_3Br$ and $PbBr_2$ were dissolved in *N*,*N*-dimethylformamide (DMF) at 0.2M independently. Then two solutions were mixed with 1:1 volume ratio and generated $CH_3NH_3Br \cdot PbBr_2$ (0.1M). The $CH_3NH_3Br \cdot PbBr_2$ solution was dip-casted onto a glass slide on a Teflon stage, which was placed in a beaker. The beaker contained 15 ml dichloromethane (DCM) leveled 3.3 cm below the Teflon stage, and was sealed by a porous Parafilm (3 M ). As DCM is a poor solution for $CH_3NH_3PbBr_3$ and lacks the ability to form hydrogen bond, the asymmetric interactions with the ions can be minimized during their assembling into crystal form. Therefore, diffusion of DCM vapor into DMF in a controlled rate can induce the growth of perovskite microstructures and nanostructures.

We first studied the synthesized perovskites with optical microscope. On the as-grown substrate, many kinds of micro and nanostructures have been observed. At the center of substrate, a lot of square shaped microdisks were well dispersed (see images and laser spectra in supplementary information). On the edge of the wafer, some microplates showed quite irregular shapes and many nanoparticles appeared. The details of perovskite nanoparticle and irregular plates can be observed clearly in the scanning electron microscope (SEM) images. Figures 1(a) and 1(c) show the SEM images of sample-A and sample-B, respectively. We can see that both of them are quite large and do not have well defined polygon shapes. Their corresponding high resolution SEM images show that sample-A (Fig. 1(b)) is primarily a thin film with irregular boundaries and randomly doped 100nm - 200 nm nanocrystals, whereas sample-B (Fig. 1(d)) is self-assembled irregular nanoparticles. Both of them can be considered as disordered perovskites.

Then the samples were excited by a frequency doubled Ti:Sapphire laser (400nm, pulse duration 100fs, and repetition rate 1kHz) and their spectra were analyzed by CCD coupled spectrometer (see Fig. 2(a) and method). Sample-A was first studied in optical characterization. When the pumping intensity was low, a broad photoluminescence peak has been observed (see solid line in Fig. 2(c)). The center wavelength was around 537nm and the full wave half maximum (FWHM) was about 25 nm. This is a very typical photoluminescence emission of $CH_3NH_3Br \cdot PbBr_2$ perovskite crystals. With the increase of pumping density, a sharp peak at 548 nm emerged in the emission spectra (see dashed line in Fig. 2(c)) and quickly dominated the emission (see dotted line in Fig. 2(c)). The high-resolution spectrum is shown as green line in Fig. 2(b), where a number of discrete lasing peaks can be resolved. The FWHM of narrowest laser spike was only around Δλ = 0.25nm. At the threshold point, the FWHM was about 1 nm, giving a Q (Q ~ λ/δλ) factor around 500. This linewidth is two orders of magnitude smaller than the photoluminescence and thus indicates the lasing actions well.

To confirm the onset of lasing actions in disordered perovskites, we have also studied the dependences of output intensity on the pumping density. As shown in Fig. 2(d), a clear "S" shaped curve can be observed in the log-log plot. The slopes of output intensities increased from around 1 at the beginning to 5.6 and finally changed back to around 1, corresponding the transition from spontaneous emission to amplified spontaneous emission and finally to the gain saturation. Then the lasing action can be confirmed from the S curve and the dramatically reduced FWHM. In Fig. 2(d), a laser threshold, which is the first kink in S curve, can be observed 60 μJ/cm$^2$. This value is comparable to the previous report in polygon shaped microdisks [6].

In additional to confirm the lasing action, it is also interesting to explore its underlying mechanism. While there are a few lasing peaks in Fig. 2(b), they are clearly not equally spaced. This is different from previous reports in polygon shaped microcavities [5, 6]. This difference is actually expected. As shown in Fig. 1(a), the perovskite doesn't have regular boundary shapes along any crystalline direction. Then both Fabry-Perot modes and multi-bouncing modes in previous reports can be excluded. From the SEM images, we also know that a number of nanoparticles and microholes randomly embedded in perovskites. These nanoparticles and microholes can function as scattering centers, which further suppress the conventional lasers and provide multiple scattering within the gain media. This can also be confirmed from the fluorescent microscope image in Fig. 2(b), where several randomly distributed strong scattering points can be clearly observed. Associated with the non-uniform fluorescence microscope image and disordered nanostructures in SEMs, the non-periodic lasing spikes of green line in Fig. 2(b) can be considered as random lasers. To give a further confirmation, we have also measured the lasing spectra as a function of pumping positions. As the orange line in Fig. 2(b), slight shift in pumping center can dramatically affect both of the peak numbers and their wavelengths, showing that the lasing actions in experiments have different light confinements and are strongly dependent on the local structure. This is also consistent with the previous experimental observations [17] and theoretically predictions of random lasers well [24].

In general, random lasers are mirrorless lasers that don't reply on the quality of cavity boundary and are very generic in many types of materials. Here random lasers have also been observed in a number of samples. Figure 3 shows the experimental results of sample-B. In 3(a), multiple sharp peaks can be observed when sample-B was excited at 87.5 µJ/cm$^2$. The mode spacings vary between 0.695 and 0.917 (see inset in Fig. 3(a)). Such a variation is much larger than conventional whispering gallery modes in regular shaped microcavities [30-32]. The random lasers can also be confirmed by the fluorescent microscope images. In additional to the cavity boundaries, there are several bright spots inside the random media (see insets in Figs. 2(b) and 3(b)). These bright spots confirm the scattering inside the random medium well. Thus the random lasing actions can be confirmed by the non-uniform fluorescent microscope image and randomly distributed laser peaks. Figure 3(b) shows the integrated intensities of outputs as a function of pumping density. Similar to sample-A, the output intensity increased dramatically when pumping density was above 75 µJ/cm$^2$. We note that the threshold of sample-B is much larger than sample-A. This is because the scattering is much stronger in sample-B, which can be seen in the SEM images in Figs. 1(b) and 1(d). Beside the increase in laser threshold, stronger scattering also forms much better light confinements and thus the corresponding laser spectrum shall be dominated by laser peaks instead of amplified spontaneous emission (ASE). This can also be found by comparing the laser spectra in Figs. 2(b) and 3(a). The laser peaks in Fig. 3(a) are more obvious.

To further confirm the formation of random lasers, we have numerically calculated the eigenfrequencies of random structures. The simulated structure was constructed by directly importing SEM images to define the outer boundary, followed by adding randomly distributed nanoparticles inside to form multiple scattering. For simplicity, here the refractive indices of perovskite nanocrystals and surrounding amorphous perovskite are 2.5 and 2.2, respectively. The material dispersion has been neglected. The outside environment is air with n = 1. Due to the dependence of emission properties on crystal quality, the perovskites nanocrysltas are considered to have rectangle shapes (can also see in SEM images, and see the supplementary inormation). Figure 4(a) illustrates the calculated Q factors of resonances inside disordered perovskite. In the lasing spectral range, more than a hundred resonances have been observed. While their Q factors vary mode-by-mode, their exact values are very close (a few hundred to thousand). This is a very general result in calculating random resonances. In real experiment, the mode number will be decreased by the nonlinear mode interactions and form a few lasing modes [24].

The field patterns of all resonances have also been studied simultaneously. As the wave localizations are primarily formed by the multiple scattering, we can see that the main fields are confined away from the cavity boundary. This is consistent with the near field microscope image (see insets in Figs. 2(b) and 3(b)), where maximum spots are randomly distributed inside the perovskites. Another important information obtained from numerical calculation is that different modes are confined within different regions. The field patterns of mode-1 and mode-2 are shown in Fig. 4(b) as examples. While their resonant frequencies are very close in Fig. 4(a), their field patterns are totally different at many regions, e.g. the bottom corners of sample in Fig. 4(b) (see more details in supplementary information). Therefore, it is easy to know that moving the pumping position in real experiment can effectively change the lasing modes (see Fig. 2(b) and supplementary information).

As we mentioned above, perovskites and random lasers have common ground in lost-cost and thus can be functioned as biological detector. For such kind of applications, the penetration depth of pumping laser is always a problem. To solve this problem, based on the high two-photon absorption coefficient [11, 12], we have also explored the possibility of constructing two-photon pumped random lasers [33, 34]. All the results are plotted in Fig. 5. When the sample-C was excited by a 800nm Ti:Sapphire laser at 3.74 mJ/cm$^2$, bright green emission could be observed in microscope. Figure 5(a) shows the recorded emission spectrum, where a broad peak can be observed around 550 nm (solid line in Fig. 5(a)). Since the frequency of pumping laser is smaller than the bandgap of perovskite ($CH_3NH_3PbBr_3$), the green light emission clearly confirmed the occurrence of two-photon absorption. When the pumping density was increased, the FWHM of photoluminescence decreased and finally spikes appeared in Fig. 5(a). The corresponding high-resolution spectrum in the inset of Fig. 5(a) shows that the relative broad peak consists of a number of narrow peaks with variable mode spacing and Q factors. These observations are consistent well with the single-photon excited random lasers. The near-field microscope image in inset-1 of Fig. 5(b) also confirms the random laser actions. Here the maxima points are distributed at the center of perovskites instead of the cavity boundary. This shows that the light is mainly confined by multiple scattering inside the perovskite instead of the reflections at the boundary. Figure 5(b) shows that the dependence of output intensity on the pumping density. A clear S shaped threshold behavior can be observed, giving a threshold around 4.4 mJ/cm$^2$. This value is about one order of magnitude higher than the thresholds under single-photon excitation and ensures the potential applications in biological detection or imaging.

## Conclusion

In summary, we have synthesized the solution-processed perovskites and studied their lasing behaviors. With self-assembled perovskite nanoparticles, we have observed random lasing actions with low threshold and relative high Q factors. These observations are attributed to the high quantum yields of perovskites and the mirrorless characteristics of random lasers. Based on the recent measured two-photon absorption coefficient of perovskites, we have also explored the lasing actions in random gain media under two-photon pumping. Since the nanoparticles (or core-shell structures) have the potentials to enter biological tissue and assemble around particular cancers. We believe that our research shall find its applications in biological detection, especially in the case that the two-photon pumped lasing actions have also been demonstrated.

## Methods:

**Optical measurement:** The perovskite samples were pumped with a regenerative amplifier seeded by a Ti:Sapphire laser (Spectra Physics, Matai, 800nm, 100fs, 86M Hz). The incident laser was frequency doubled by a BBO crystal and coupled to a home-made microscope system and focused by a 60x objective lens (NA = 0.7). The emitted lights were collected and collimated by the same objective lens and coupled to a CCD (Princeton Instrument, Pixis CCD) coupled spectrometer (Princeton Instrument, Spectropro SP2700i) via a multimode fiber. The near-field image under optical excitation was taken by the same microscope.

**Numerical calculations:** Based on the SEM image of perovskite nanoparticles, we have numerically studied their resonant properties with finite element method (Comsol multiphysics). In our numerical calculation, the three-dimensional structures were simplified into a simple two-dimensional object by using effective refractive index. The real structure was obtained by directly importing the SEM picture. By using the perfect matching layer to absorb the outgoing waves, complex valued eigenfrequencies ($\omega$) can be obtained. The Q factors are determined by $Q = Re(\omega)/2|Im(\omega)|$.

## Acknowledgement:

This work is supported by NSFC11204055, NSFC61222507, NSFC11374078, NCET-11-0809, Shenzhen Peacock plan under the Nos. KQCX20120807091143322 and KQCX20130627094615410, and Shenzhen Fundamental research projects under the Nos. JCYJ20130329155148184, JCYJ20140417172417110, JCYJ20140417172417096.

## Author contributions:

## Additional Information

**Competing financial interests:** The authors declare no competing financial interests.

# Figure Captions:

**Fig. 1**: (a) (c) SEM images of sample-1 and sample-2. (b) and (d) The high resolution SEM images of sample-1 and sample-2 to show the detail nanostructures.

**Fig. 2**: (a) Schematic picture of the setup for optical measurement. (b) Laser spectra at different positions and the near field microscope image of sample-1. Here the pumping density is 75 μJ/cm$^2$. (c) Low-resolution of laser spectra under different pumping density. (d) Integrated output intensity as a function of pumping density. A threshold can be observed at 60 μJ/cm$^2$.

**Fig. 3**: (a) Laser spectrum of sample-B under optical excitation at 87.5 μJ/cm$^2$. Inset shows the variation of mode spacings. (b) The dependence of output intensity on the pumping density. Inset is the corresponding near field microscope image.

**Fig. 4**: (a) Numerically calculated Q factors of resonances inside the random medium. (b) The calculated near field patterns of mode-1 and mode-2 in (a). While their resonant wavelengths are very close, the corresponding field patterns are quite different.

**Fig. 5**: (a) The spectra of emissions from perovskite under two-photon excitation with different pumping density. Inset shows the high-resolution spectrum at 6.63 mJ/cm$^2$. (b) The dependence of output intensity on the pumping density. Inset-I and inset-II show the corresponding near field image and SEM image of perovskite.

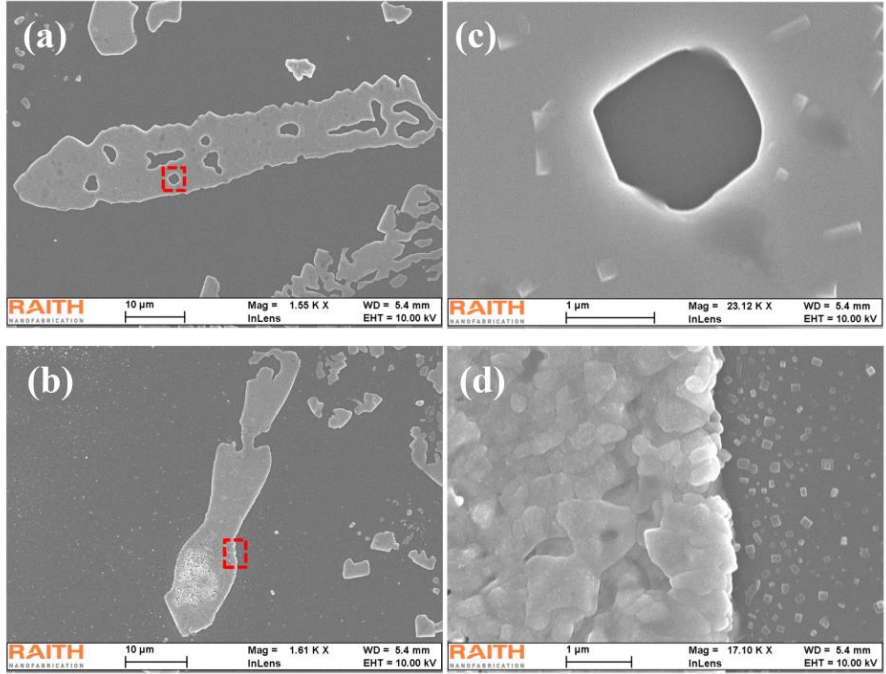

Fig. 1

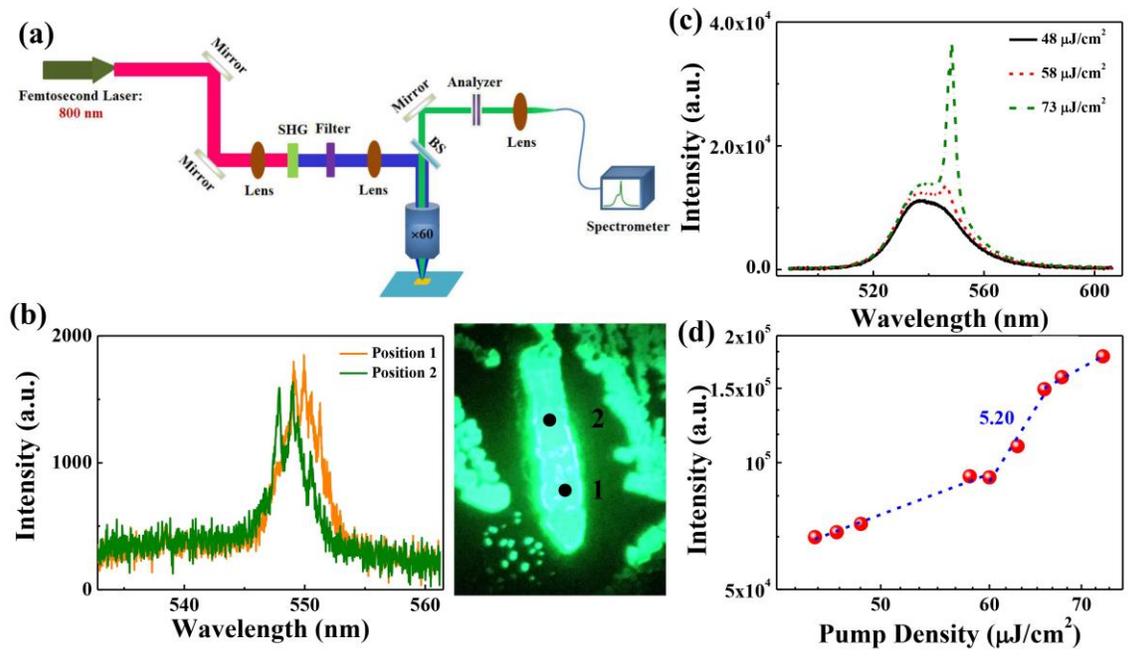

Fig. 2

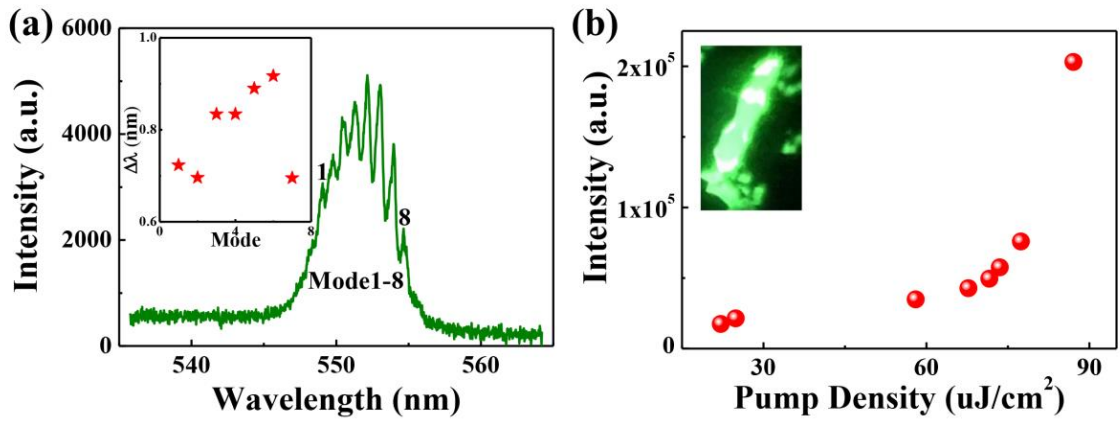

Fig. 3

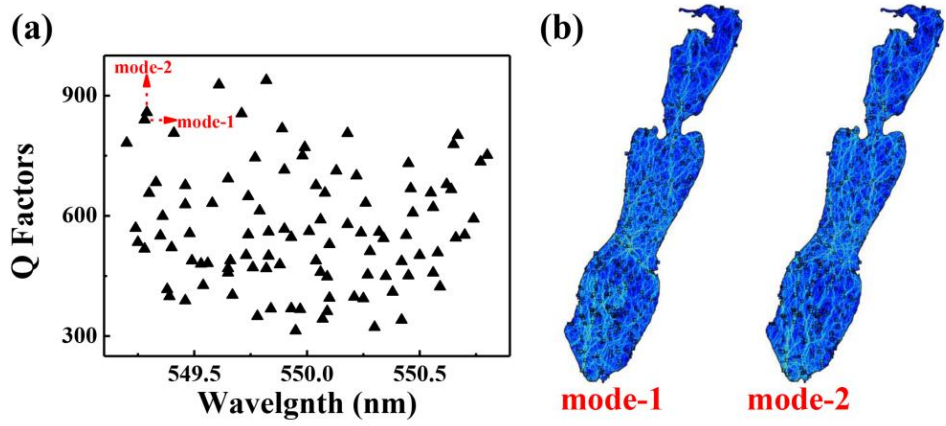

Fig. 4

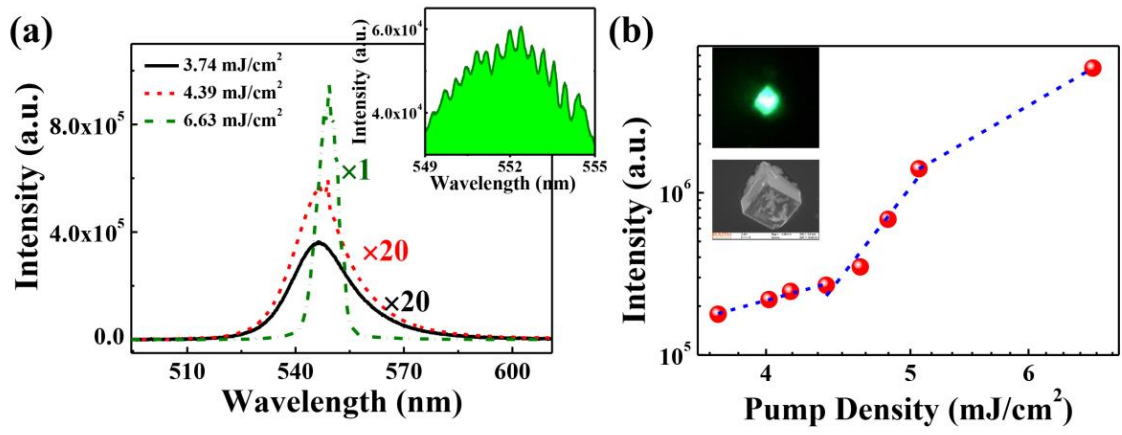

Fig. 5